\documentclass{article}
\usepackage{spconf,amsmath,graphicx,subfigure,enumitem,cleveref}
\usepackage[colorlinks=true]{hyperref}
\usepackage{booktabs,multirow}
\usepackage[normalem]{ulem}
\usepackage{color,soul}
\useunder{\uline}{\ul}{}

\title{Sample-level CNN Architectures for Music Auto-tagging \\ Using Raw Waveforms}
\name{Taejun Kim$^{1}$,
      Jongpil Lee$^{2}$,
      Juhan Nam$^{2}$}

\address{$^1$ School of Electrical and Computer Engineering, University of Seoul,\\
         $^2$ Graduate School of Culture Technology, KAIST,\\
         ktj7147@uos.ac.kr,
         \{richter, juhannam\}@kaist.ac.kr\\
}
\begin{document}
%\ninept
%
\maketitle
\begin{abstract}
%Recent work has shown that the end-to-end approach also works well for music signals. However, 1D convolutional neural network (CNN) architectures for music auto-tagging are not fully explored. In this paper, we adopt building blocks from state-of-the-art image classification models (\textit{ResNets} and \textit{SENets}) to 1D CNNs, and show that multi-level feature aggregation is effective. As a result, our models obtain significant improvements over previous \textit{state-of-the-art} on MagnaTagATune dataset and show comparable results on Million Song Dataset. Furthermore, we analyze, and visualize our model to show how 1D CNN operates. Code and models are available at \url{https://github.com/tae-jun/resemul}.
Recent work has shown that the end-to-end approach using convolutional neural network (CNN) is effective in various types of machine learning tasks. For audio signals, the approach takes raw waveforms as input using an 1-D convolution layer. In this paper, we improve the 1-D CNN architecture for music auto-tagging by adopting building blocks from state-of-the-art image classification models, \textit{ResNets} and \textit{SENets}, and adding multi-level feature aggregation to it. We compare different combinations of the modules in building CNN architectures. The results show that they achieve significant improvements over previous state-of-the-art models on the MagnaTagATune dataset and comparable results on Million Song Dataset. Furthermore, we analyze and visualize our model to show how the 1-D CNN operates.
\end{abstract}
%

% ------------------------------------
% --- SECTION 1: INTRODUCTION
% ------------------------------------
\section{Introduction}
\label{sec:intro}
Time-frequency representations based on short-time Fourier transform, often scaled in a log-like frequency such as mel-spectrogram, are the most common choice of input in the majority of state-of-the-art music  classification algorithms \cite{choi2017convolutional,lee2017multi,dieleman2013multiscale,van2014transfer,choi2016automatic}. The 2-dimentional input represents  acoustically meaningful patterns well but requires a set of parameters, such as window size/type and hop size, which may have different optimal settings depending on the type of input signals.  

% In music information retrieval (MIR) tasks, the most systems transform 1D acoustic signals into 2D time-frequency representations based on the Fourier transform (e.g. mel-spectrograms) and then use them as input to the systems. \cite{choi2017convolutional,lee2017multi,dieleman2013multiscale,van2014transfer,choi2016automatic} Such transforms can extract acoustic features well, but they loss some features like phases. In addition, the 2D representations have a trade-off of time and frequency resolution. Therefore, it is hard to set proper hyperparameters required for the transforms (e.g. window size, hop size, etc.).

In order to overcome the problem, there have been some efforts to directly use raw waveforms as input particularly for convolutional neural networks (CNN) based models \cite{dieleman2014end,dai2017very}. While they show promising results, the models used large filters, expecting them to replace the Fourier transform. Recently, Lee et. al. \cite{lee2017sample} addressed the problem using very small filters and successfully applied the 1D CNN to the music auto-tagging task. Inspired from the well-known \textit{VGG} net that uses very small size of filters such as $3 \times 3$, \cite{simonyan2014very}, the sample-level CNN model was configured to take raw waveforms as input and have filters with such small granularity.    

% In order to overcome these limitations and further improve a performance of the systems, there were researches which apply convolutional neural networks (CNNs) directly on raw waveforms \cite{dieleman2014end,dai2017very}. However, they used large filters, expecting them to replace the Fourier transform and extract more complex features. Recently, lee et al. \cite{lee2017sample} addressed the problem and, using small filters, successfully applied 1D CNNs to music auto-tagging using raw waveforms. The models are called \textit{SampleCNNs} and have a simple CNN architecture which is just a sequence of convolutional and pooling layers like \textit{VGG} \cite{simonyan2014very}, which is a well-known image classification model.

A number of techniques to further improve performances of CNNs have appeared recently in image domain. He et. al. introduced \textit{ResNets} which includes \textit{skip connections} that enables a very deep CNN to be effectively trained and makes gradient propagation fluent \cite{he2016deep}. Using the skip connections, they could successfully train a 1001-layer ResNet \cite{he2016identity}. Hu et. al proposed \textit{SENets} \cite{hu2017squeeze} which includes a building block called \textit{Squeeze-and-Excitation} (SE). Unlike other recent approaches, the block concentrates on channel-wise information, not spatial. The SE block adaptively recalibrates feature maps using a channel-wise operation. Most of the techniques were developed in the field of computer vision but they are not fully adopted for music classification tasks. Although there were a few approaches to readily apply them to audio domain  \cite{dai2017very,hershey2017cnn}. They used 2D representations as input \cite{hershey2017cnn} or used large filters for the first 1D convolutional layer \cite{dai2017very}.

On the other hand, some methods are concerned with overall architecture of the model rather than designing a fine-grained building block \cite{lee2017multi,lee2017samplemulti,sun2014deep,donahue2014decaf,aytar2016soundnet}. Specifically, multi-level feature aggregation combines several hidden layer representations for final prediction \cite{lee2017multi,lee2017samplemulti}. They significantly improved the performance in music auto-tagging by taking different levels of abstractions of tag labels into account. 

In this paper, we explore the building blocks of advanced CNN architectures, \textit{ResNets} and \textit{SENets}, based on the sample-level CNN for music auto-tagging. Also, we observe how the multi-level feature aggregation affects the performance. The results show that they achieve significant improvements over previous state-of-the-art models on the MagnaTagATune dataset and comparable results on Million Song Dataset. Furthermore, we analyze and visualize our model built with the SE blocks to show how the 1D CNN operates. The results show that the input signals are processed in a different manner depending on the level of layers. 

\begin{figure*}[!t]
  \centering
  % Overview
  \subfigure[Overview of the architecture]{\includegraphics[height=8cm,keepaspectratio]{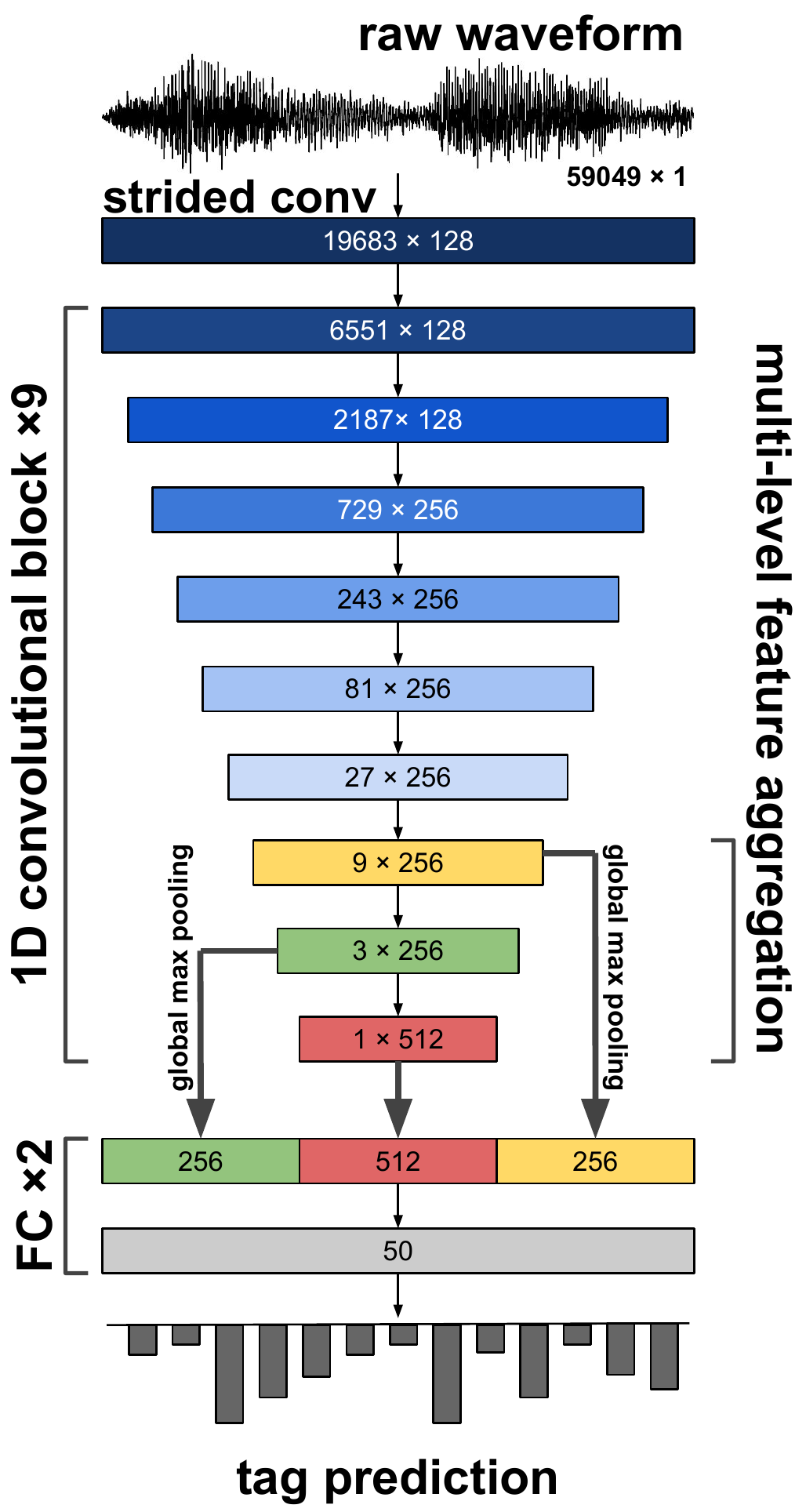}\label{fig:1a}}
  % Basic block
  \subfigure[Basic block \cite{lee2017sample}]{\includegraphics[height=8cm,keepaspectratio]{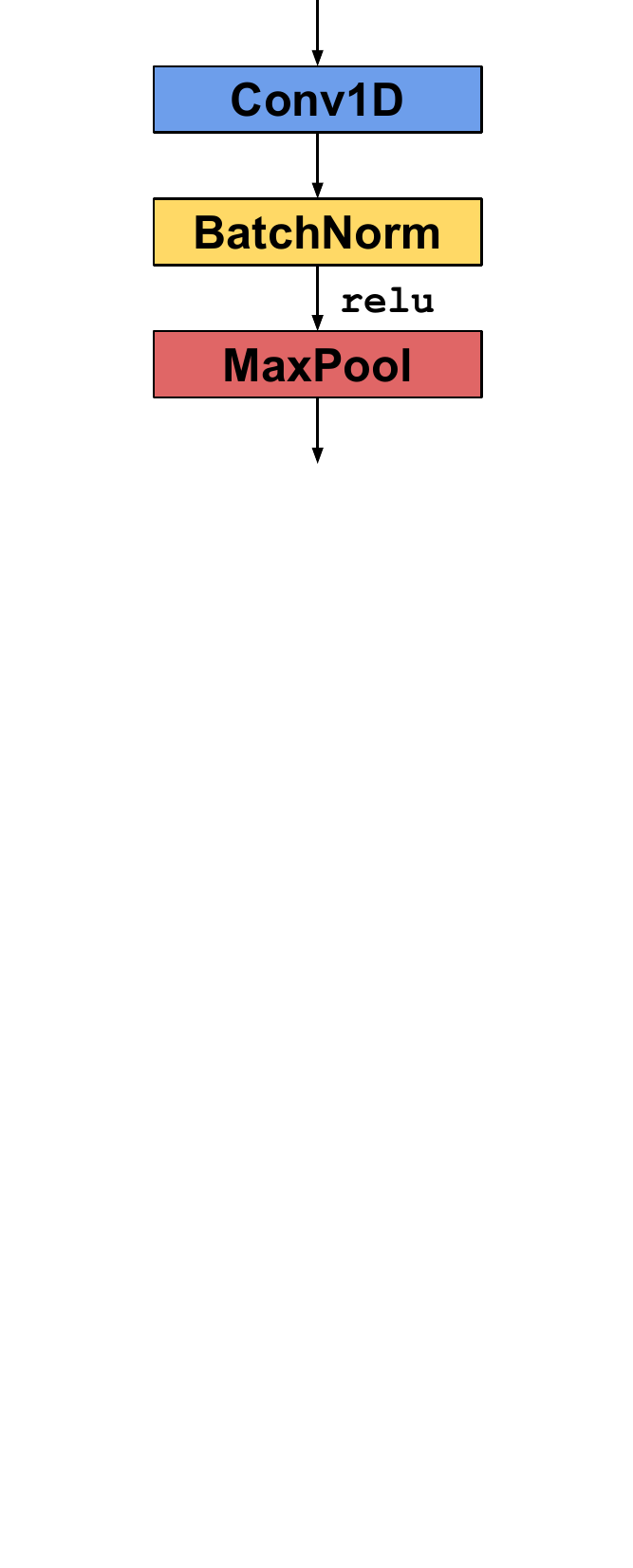}\label{fig:1b}}
  % SE block
  \subfigure[SE block]{\includegraphics[height=8cm,keepaspectratio]{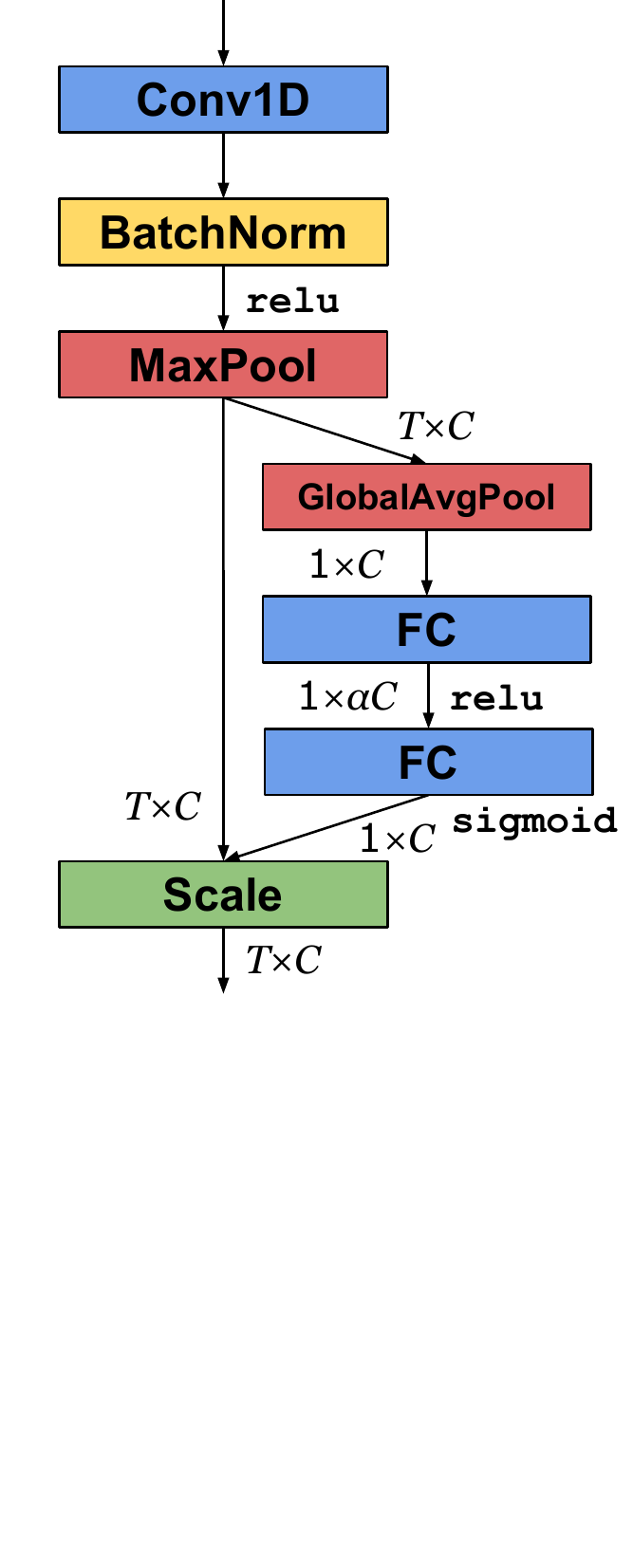}\label{fig:1c}}
  % Res-n block
  \subfigure[Res-$n$ block]{\includegraphics[height=8cm,keepaspectratio]{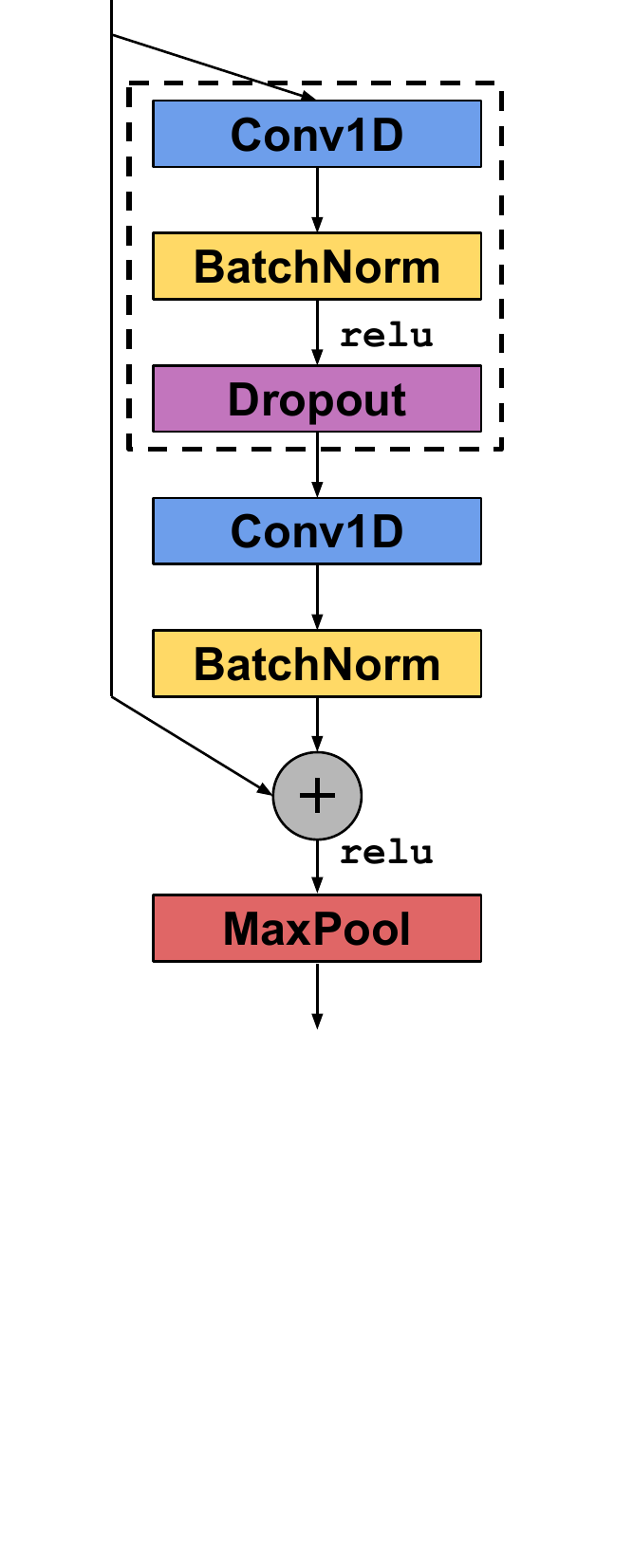}\label{fig:1d}}
  % ReSE-n block
  \subfigure[ReSE-$n$ block]{\includegraphics[height=8cm,keepaspectratio]{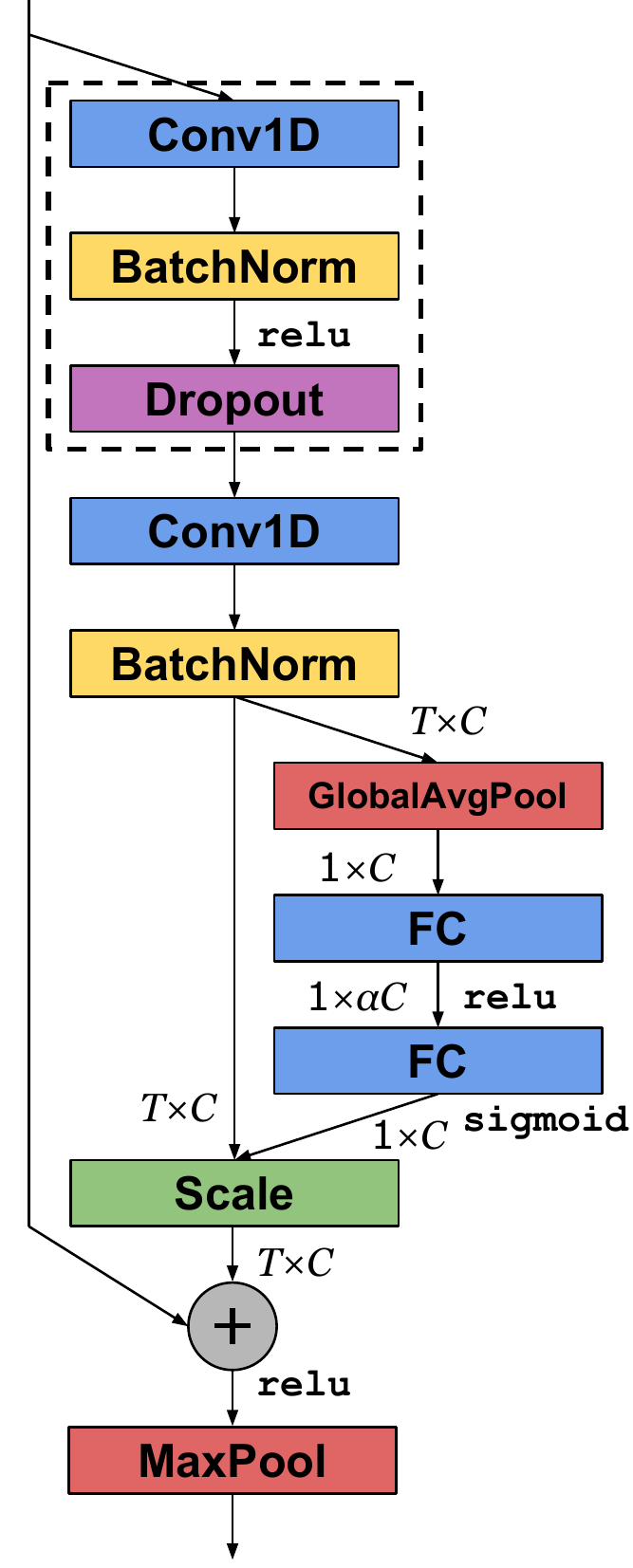}\label{fig:1e}}
  % Caption
  \caption{The proposed architecture for music auto-tagging. (a) The models consist of a strided convolutional layer, 9 blocks, and two fully-connected (FC) layers. The outputs of the last three blocks are concatenated and then used as input of the last two FC layers. Output dimensions of each block (or layer) are denoted inside of them (temporal$\times$channel). (b-e) The 1D convolutional building blocks that we evaluate.}
  \label{fig:fig1}
\end{figure*}
% (END) --- FIGURE 1: Architecture.

% ------------------------------------
% --- SECTION 2: ARCHITECTURES
% ------------------------------------
\section{Architectures}
\label{sec:arch}

All of our models are based on the sample-level 1D CNN model \cite{lee2017sample}, which is constructed with the basic block shown in Figure \ref{fig:1b}. Every filter size of the convolution layers is fixed to three. The differences between the sample-level CNN and ours are the use of advanced building blocks and multi-level feature aggregation. In this section, we describe the details.

% --- SUBSECTION 2.1: Building blocks.
\subsection{1D convolutional building blocks}
\label{ssec:blocks}

\subsubsection{SE block}
We utilize the \textit{SE} block from \textit{SENets} to increase representational power of the basic block. As shown in Figure \ref{fig:1c}, we simply attached the SE block to the basic block. The SE block recalibrates feature maps from the basic block through two operations. One is \textit{squeeze} operation that aggregates a global temporal information into channel-wise statistics using global average pooling. The operation reduces the temporal dimensionality ($T$) to one, averaging outputs from each channel. The other is \textit{excitation} operation that adaptively recalibrates feature maps of each channel using the channel-wise statistics from the squeeze operation and a simple gating mechanism. The gating mechanism consists of two fully-connected (FC) layers that compute nonlinear interactions among channels. Finally, the original outputs from the basic block are rescaled by channel-wise multiplication between the feature map and the sigmoid activation of the second FC layer.

%We utilize the \textit{SE} block of \textit{SENets} to increase representational power of the basic block. We just attach the SE block after the basic block. The SE block recalibrates feature maps from the basic block through two operations: 1) \textit{squeeze} operation aggregates a global temporal information into channel-wise statistics using global average pooling. The operation shrink a temporal dimensionality ($T$) to one, averaging outputs along each channel. 2) Accepting the channel-wise statistics from the squeeze operation, \textit{excitation} operation adaptively recalibrates feature maps of each channel, using a simple gating mechanism. The gating mechanism consists of two fully-connected (FC) layers. After the layers computing a nonlinear interactions among channels, the original outputs from the basic block are rescaled by channel-wise multiplication between the feature map and the sigmoid activation of the second FC layer.

Unlike the original SE block in \textit{SENets}, our excitation  operation does not form a bottleneck. On the contrary, we expand the channel dimensionality ($C$) to $\alpha C$ at the first FC layer, and then reduce the dimensionality back to $C$ at the second layer. We set the amplifying ratio $\alpha$ to be 16, after a grid search with $\alpha=[2^{-3}, 2^{-2},..., 2^{6}]$.

\subsubsection{Res-n block}
Inspired by \textit{skip connections} from \textit{ResNets}, we modified the basic block by adding a skip connection as shown in Figure \ref{fig:1d}. \textit{Res-$n$} denotes that the block uses $n$ convolutional layers where $n$ is one or two. Specifically, \textit{Res-2} is a block that has the additional layers denoted by the dotted line in Figure \ref{fig:1d}, and \textit{Res-1} is a block that has a skip connection only. When the block uses two convolutional layers (\textit{Res-2}), we add a dropout layer (with a drop ratio of 0.2) between two convolutions to avoid overfitting. This technique was firstly introduced at \textit{WideResNets} \cite{zagoruyko2016wide}.

\subsubsection{ReSE-n block}
The \textit{ReSE-$n$} block is a combination of the SE and Res-$n$ blocks as shown in Figure \ref{fig:1e}. $n$ denotes the number of convolutional layers in the block, where $n$ is also one or two. A dropout layer is inserted when $n$ is two.
% (END) --- SUBSECTION 2.1: Building blocks.

% --- SUBSECTION 2.2: Multi-level feature aggregation.
\subsection{Multi-level feature aggregation}
\label{ssec:multi}
Fig. \ref{fig:1a} shows the multi-level feature aggregations that we configured. The outputs of the last three blocks are concatenated and then delivered to the FC layers. Before the concatenation, temporal dimensions of the outputs are reduced to one by a global max pooling. Unlike \cite{lee2017multi}, the concatenation occurs while training the CNN and the average pooling over the whole audio clip (i.e. 29 second long), which followed by the global max pooling, is not included.  

% (END) --- SUBSECTION 2.2: Multi-level feature aggregation.

% ------------------------------------
% (END) --- SECTION 2: ARCHITECTURES
% ------------------------------------

% ------------------------------------
% --- SECTION 3: Experiments
% ------------------------------------
\section{Experiments}
\label{sec:exp}

\subsection{Datasets}
We evaluated the proposed architectures on two datasets, MagnaTagATune (MTAT) dataset \cite{law2009evaluation} and Million Song Dataset (MSD) annotated with the Last.FM tags \cite{bertin2011million}. We split and filtered both of the datasets, following the previous work \cite{choi2016automatic,dieleman2014end,lee2017sample}. We used the 50 most frequent tags. All songs are trimmed to 29 seconds long, and resampled to 22050Hz as needed. The song is divided into 10 segments of 59049 samples. To evaluate the performance of music auto-tagging which is a multi-class and multi-label classification task, we computed the Area Under the Receiver Operating Characteristic curve (AUC) for each tag and computed the average across all 50 tags. During the evaluation, we average predictions across all segments.

% 태그기반 검색 task에서의 성능 측정을 위하여 tag별로 auc를 구한후 이를 평균한 값을 평가 메트릭으로 사용한다. multi-label classification Task인 음악오토태깅에서 태그기반 검색에서의 성능 측정을 위하여 tag별로 auc를 구한후 이를 평균한 값을 평가 메트릭으로 사용한다.

% --- TABLE 1: MTAT building block comparison.
\begin{table}[t]
\centering
\caption{AUCs of CNN architectures on MTAT. ``multi'' and ``no multi'' indicates if the multi-level feature aggregation is used or not. $\dagger$ denotes using a weight decay of $10^{-4}$.}
\label{tab:arch}
\vspace{1.5mm}
\begin{tabular}{l@{\qquad}cc}
\toprule
\multirow{2}{*}{\raisebox{-\heavyrulewidth}{Block}} & \multicolumn{2}{c}{MTAT} \\
\cmidrule{2-3}
& multi & no multi  \\
\midrule
Basic \cite{lee2017sample} & 0.9077                     & 0.9055              \\
SE                         & \textbf{0.9111}            & 0.9083              \\
Res-1                      & 0.9037                     & 0.9048              \\
Res-2                      & 0.9098                     & 0.9061              \\
ReSE-1                     & 0.9053                     & 0.9066              \\
ReSE-2                    & \textbf{0.9113}$^{\dagger}$ & 0.9102$^{\dagger}$  \\
\bottomrule
\end{tabular}
\end{table}
% (END) --- TABLE 1: MTAT building block comparison.

\subsection{Implementation details}
All the networks were trained using SGD with Nesterov momentum of 0.9 and mini-batch size 23. The initial learning rate is set to 0.01, decayed by a factor of 5 when a validation loss is on a plateau. None of the regularizations are used on MSD. A dropout layer of 0.5 was inserted before the last FC layer on MTAT. For all building blocks, we evaluated either with or without the multi-level feature aggregation. Since the training for MSD takes much time longer than MTAT, we explored the architectures mainly on MTAT, and then trained the two best models on MSD. Code and models built with TensorFlow and Keras are available at the link\footnote{\url{https://github.com/tae-jun/resemul}}.

% --- TABLE 2: SoTA comparison.
\begin{table}[t]
\centering
\caption{AUCs of state-of-the-art models on MTAT and MSD. $\dagger$ denotes that the model used an ensemble of three.}
\label{tab:sota}
\vspace{1.5mm}
\begin{tabular}{l|cc}
\hline
Model                                                       & MTAT                & MSD                          \\ \hline
Bag of multi-scaled features \cite{dieleman2013multiscale}  & 0.8980              & -                            \\
End-to-end \cite{dieleman2014end}                           & 0.8815              & -                            \\
Transfer learning \cite{van2014transfer}                    & 0.8800              & -                            \\
Persistent CNN \cite{liu2016applying}                       & 0.9013              & -                            \\
Time-Frequency CNN \cite{gucclu2016brains}                  & 0.9007              & -                            \\
Timbre CNN \cite{pons2017timbre}                            & 0.8930              & -                            \\
2D CNN \cite{choi2016automatic}                             & 0.8940              & 0.8510                       \\
CRNN \cite{choi2017convolutional}                           & -                   & 0.8620                       \\
Multi-level \& multi-scale \cite{lee2017multi}              & 0.9017$^{\dagger}$  & \textbf{0.8878}$^{\dagger}$  \\
SampleCNN multi-features \cite{lee2017samplemulti}          & 0.9064$^{\dagger}$  & 0.8842                       \\
SampleCNN \cite{lee2017sample}                              & 0.9055              & 0.8812                       \\ \hline
SE [This work]                                              & \textbf{0.9111}     & 0.8840                       \\
ReSE [This work]                                            & \textbf{0.9113}     & 0.8847                       \\ \hline
\end{tabular}
\end{table}
% (END) --- TABLE 2: SoTA comparison.

% --- FIGURE 2: Excitation viz.
\begin{figure}[t]
  \centering
  \includegraphics[width=1.0\linewidth]{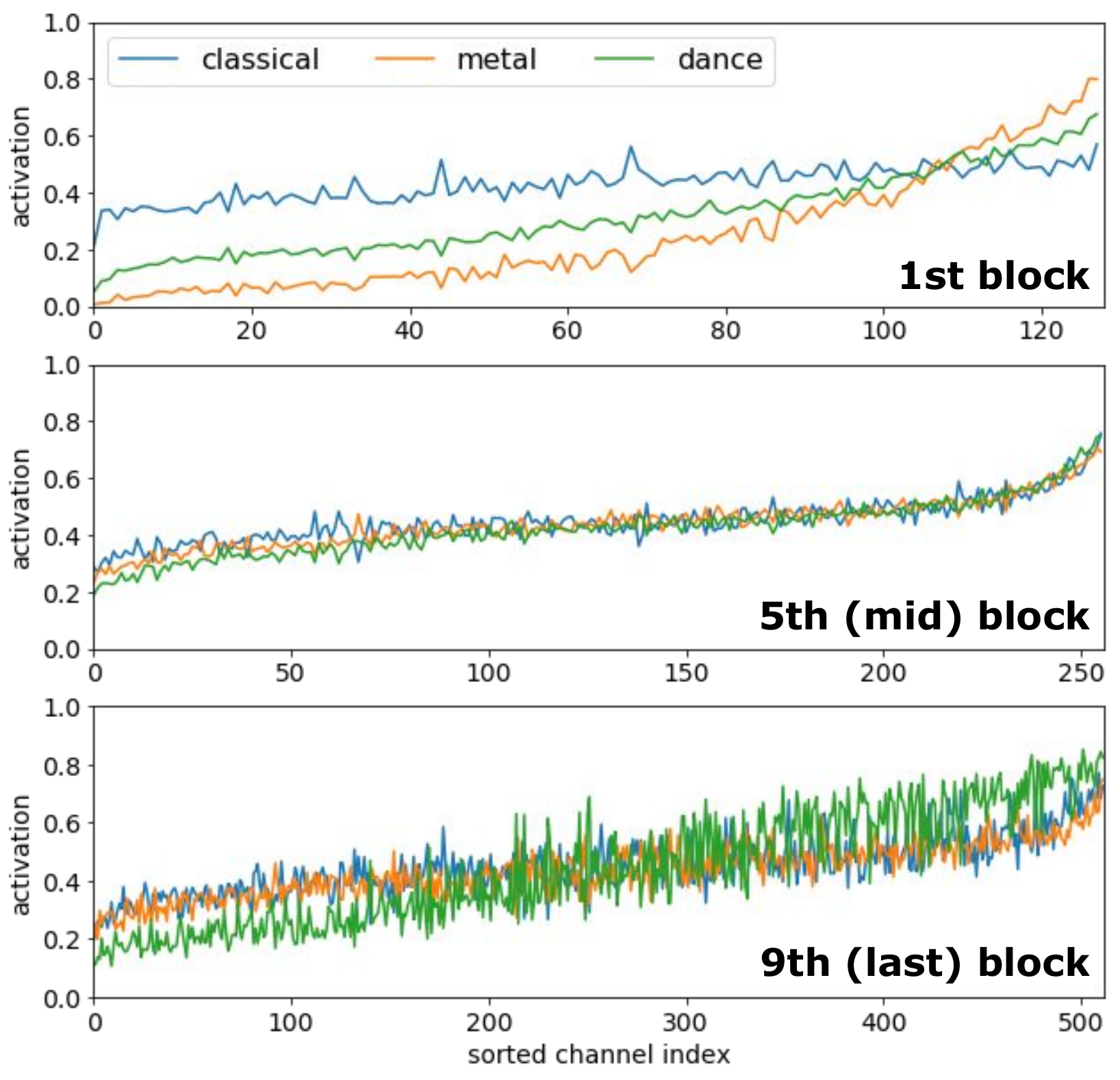}
  \caption{Visualization of the sigmoid activations of \textit{excitations} in the \textit{SE} model. The channel index was sorted by the average of the activations.}
  \label{fig:exviz}
\end{figure}
% (END) --- FIGURE 2: Excitation viz.

\section{Results and Discussion}

\subsection{Comparison of the architectures}
Table \ref{tab:arch} summarizes the evaluation results of compared CNN architectures on the MTAT dataset. They show that the SE block is more effective than the Res-$n$ blocks, increasing the performance of the basic block for all cases. In the Res-$n$ block, only adding the skip connection to the basic block (Res-1) actually decreases the performance. The combination of the SE and the Res-2 improves it slightly more. However, a training time of the ReSE-2 is 1.8 times longer than the basic block whereas the SE block only 1.08 times longer. Thus, if the training or prediction time of the models is important, the SE model will be preferred to the ReSE-2. The effect of the multi-level aggregation is valid for the majority of the models. We obtained two best results in Table \ref{tab:arch} by using the multi-level aggregation. 

% \textbf{The building blocks}. With a slight change, the SE block increases the performance of the basic block significantly. Only adding the skip connection to the basic block (Res-1) decreases the performances. But, it was effective with an additional convolutional layer (Res-2). The combination of the SE and the Res-$n$ were powerful (ReSE-$n$). However, a training time of the ReSE-2 is 1.8 times longer than the basic block, whereas the SE block only 1.08 times longer. If a training time or a prediction speed of a model is important, then SE model might be preferred than ReSE-2, even though the performance of ReSE-2 is higher.

% \textbf{The multi-level feature aggregation}. The effect of the multi-level aggregation is valid for majority of the models. The both of two highest models are used the aggregation.

\subsection{Comparison with state-of-the-arts}
Table \ref{tab:sota} compares previous state-of-the-art models in music auto-tagging with our best models, the SE block and ReSE-2 block, each with multi-level aggregation. On the MTAT dataset, our best models outperform all the previous results. On MSD, they are not the best but are comparable to the second-tier.

\section{Analysis of Excitation}
\label{ssec:exviz}
To lay the groundwork for understanding how 1D CNNs operate, we analyze the sigmoid activations of \textit{excitations} in the SE blocks at different levels graphically and quantitatively. In this section, we observe how the SE blocks recalibrate channels, depending on which level they exist.  The blocks used for the analysis are from the SE model using the multi-level feature aggregation and they were trained on MTAT. The activations were extracted from its test set. The activations were averaged over all segments separately for each tag.

%\textbf{Graphical analysis}.
\subsection{Graphical analysis}
For this analysis, we chose three tags, \textit{classical}, \textit{metal}, and \textit{dance} that are not similar to each other as shown in Table \ref{tab:cooc}. Figure  \ref{fig:exviz} shows the average sigmoid activations in the SE blocks for the songs with the three tags. The different levels of activations indicate that the SE blocks process input audio differently depending on the tag (or genre) of the music. That is, every block in Figure \ref{fig:exviz} fires different patterns of activations for each tag at a specific channel. This trend is strongest at the first block (top), weakest at the mid block (middle), and becomes stronger again at the last block (bottom). %This is observed in Figure \ref{fig:exstd} where the standard deviation of the activations for all tags are measured. 

% While the exclusive trend is strong at the first and last block, their patterns are quite different. The activations of the first block are consistent for each tag. Specifically, the first block usually fires high activations for \textit{classical}, low ones for \textit{dance}, and even lower ones for \textit{metal}. On the other hand, the activations of the last block vary depending on the tags. Specifically, the activations of \textit{metal} are high at some channels but low at the others, which makes the activations noisy even though they are sorted. 

This trend is somewhat different from what are observed in the image domain \cite{hu2017squeeze}, where the exclusiveness of average excitation for input with different labels are monotonically increasing along the layers. Specifically, the first block fires high activations for \textit{classical}, low ones for \textit{dance}, and even lower ones for \textit{metal} for the majority of the channels. On the other hand, the activations of the last block vary depending on the tags. For example, the activations of \textit{metal} are high at some channels but low at the others, which makes the activations noisy even though they are sorted. We can interpret this result as follows. The first block normalizes the loudness of the audios because the block fires high activations for \textit{classical} music, which tend to have small volume, and low activations for \textit{metal} music, which tend to have large volume. Also, the middle block processes common features among them as they have similar levels of activations. Finally, the noisy exclusiveness in the last block indicates that they effectively discriminate the music with different tags.

% In conclusion, we interpret these phenomena as the first block normalizing volumes of the audios, since the block fires high activations for \textit{classical} audios (small volume), and low activations for \textit{metal} audios (big volume). And, we interpret the mid block as generally processing the signal, and the last block discriminatively processing the signal.

% The exclusive trend is strong at the first and last block, however, there is a huge difference between their patterns. The activations of the first block are consistent for each tag, but not of the last. More specifically, the first block usually fires high activations for \textit{classical}, low ones for \textit{dance}, and even lower ones for \textit{metal}. But, the last block fires inconsistent activations. The activations of the last block vary depending on the tags, but the activations of \textit{metal} are high at some channels and low at the others, which makes the activations noisy even they are sorted. In conclusion, we interpret these phenomena as the first block normalizing a signal, the mid block generally processing the signal, and the last block discriminatively processing the signal.

%\textbf{Quantitative analysis}. 
\subsection{Quantitative analysis}
\begin{figure}[t]
  \centering
  \includegraphics[width=0.7\linewidth]{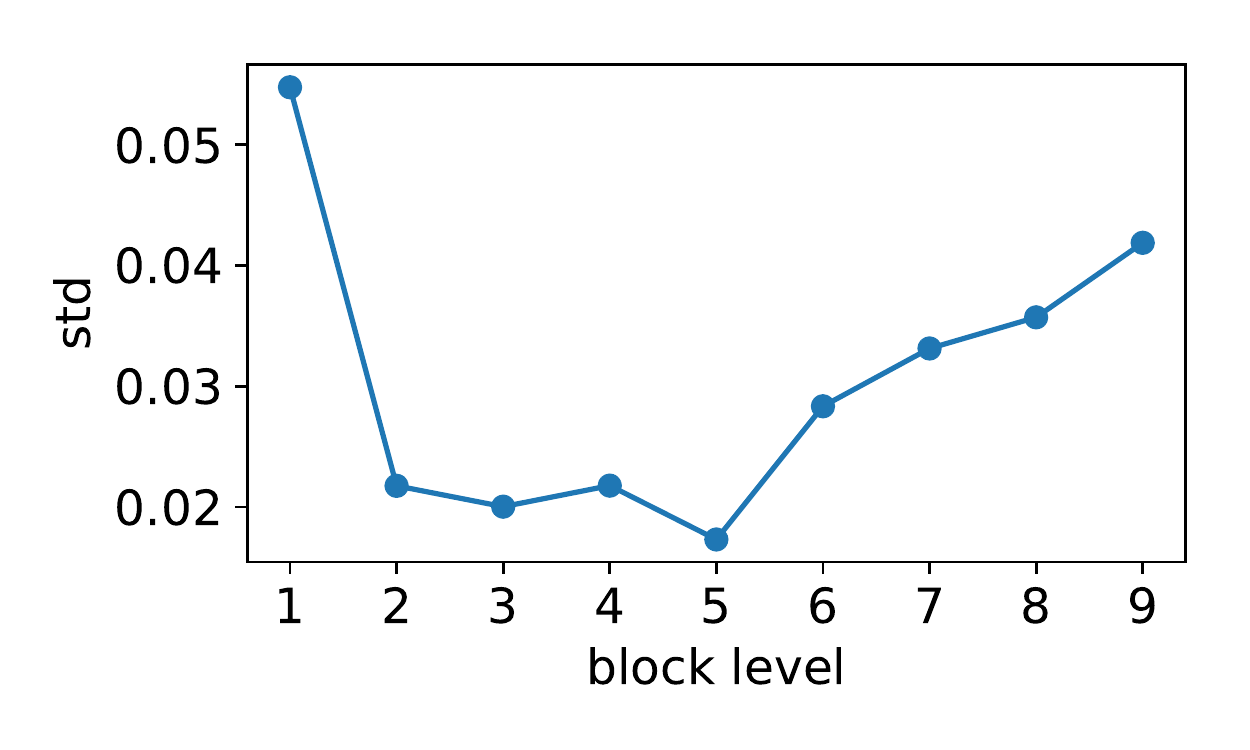}
\vspace{-2mm}
\caption{Standard deviations (std) of the activations of \textit{excitations} across all tags along each layer.}
  \label{fig:exstd}
\end{figure}

We assure the exclusiveness trend by measuring standard deviations of the activations across all tags at every level. Figure \ref{fig:exstd} shows that the higher the standard deviation is, the more the block responses to the song differently according to its tag. The result shows that the standard deviation is highest at the first block, it drops and stays low up to the 5th block and then increases gradually until the last block. That is, the four lower blocks except the the bottom one (2 to 5) tend to handle general features whereas the four upper blocks (6 to 9) tend to progressively more discriminative features.

% ------------------------------------
% (END) --- SECTION 3: Experiments
% ------------------------------------

% --- TABLE 3: Co-occurrence matrix
\begin{table}[t]
\centering
\caption{Co-occurrence matrix of the tags used in Figure  \ref{fig:exviz}}
\label{tab:cooc}
%\vspace{1mm}
\begin{tabular}{c||c|c|c}
\toprule
                   & \textit{classical} & \textit{metal} & \textit{dance} \\ \hline \hline
\textit{classical} & 704               & \textbf{0}     & \textbf{1}     \\ \hline
\textit{metal}     & \textbf{0}         & 166           & \textbf{0}     \\ \hline
\textit{dance}     & \textbf{1}         & \textbf{0}     & 153           \\ \bottomrule
\end{tabular}
\end{table}
% (END) --- TABLE 3: Co-occurrence matrix

% --- FIGURE 3: Excitation std.
% (END) --- FIGURE 3: Excitation std.

% ------------------------------------
% --- SECTION 4: Conclusion
% ------------------------------------
\section{Conclusion}
\label{sec:conc}
We proposed 1D convolutional building blocks based on the previous work, the \textit{sample-level CNN}, \textit{ResNets}, and \textit{SENets}. The \textit{ReSE} block, which is a combination of the three models, showed the best performance. Also, the multi-level feature aggregation showed improvements on the majority of the building blocks. Through the experiments, we obtained state-of-the-art performance on the MTAT dataset and high-ranked results on MSD. In addition, we analyzed the activations of \textit{excitation} in \textit{SE} model to understand the effect. With this analysis, we could observe that the SE blocks process non-similar songs exclusively and how the different levels of the model process the songs in a different manner.   

% an audio processed in the model- normalization at the beginning, general processing at the middle, and discriminate at the end. 
% ------------------------------------
% (END) --- SECTION 4: Conclusion
% ------------------------------------

\vfill\pagebreak

\bibliographystyle{IEEEbib}
\bibliography{refs}

\end{document}